\documentstyle[12pt, aaspp4]{article}

\begin{document}

\title{The gravitational lens MG0414+0534: a link between red galaxies and
dust
\footnote{The observations reported here were obtained at the Multiple
Mirror Telescope, a joint facility of the University of Arizona and the
Smithsonian Institution.}}

\author{B. A. McLeod}
\affil{Harvard-Smithsonian Center for Astrophysics, 60 Garden St.,
Cambridge, MA 02138}

\author{G. M. Bernstein}
\affil{University of Michigan, Department of Astronomy, 830 Dennison Building,
Ann Arbor, MI 48109}

\author{M. J. Rieke}
\affil{University of Arizona, Steward Observatory, Tucson, AZ 85721}

\and

\author{D. W. Weedman}
\affil{Pennsylvania State University, Department of Astronomy \& Astrophysics, 
525 Davey Lab., University Park, PA 16802}

\begin{abstract}
We present near infrared observations of the red gravitational lens
system MG0414+0534.  Our images are of sufficient quality to allow
photometry of all four QSO images and the lens galaxy. The
observations show that the $K$-band brightnesses of the components are
more similar to the radio brightnesses than to the optical and thus
support the notion that the system is highly reddened.  The differing
brightnesses at visible wavelengths are interpreted as variable
amounts of extinction in the lens galaxy. The lens galaxy is
detected at $K$-band and is as red as any other known galaxy of
comparable apparent magnitude.  An estimate of the redshift of the
lens galaxy of $0.45 < z < 0.6$ is determined from the
Faber-Jackson relation and photometric considerations.  By extension
we argue that other very red field galaxies may contain large amounts
of dust.  $K$-band spectra of the individual components show no
difference in the H$\alpha$ equivalent widths among the components.
This is evidence against significant microlensing.

\end{abstract}

\section{Introduction}

Gravitational lensing of quasars by foreground galaxies provides us
with a probe into the universe at high redshift.  Useful information
may be gleaned from lensing systems including limits on the
cosmological constant and the properties of high-redshift lens
galaxies.  Lensing surveys also have the potential for detecting
massive objects in the universe that might not otherwise have been
located. Radio surveys are particularly important for studying the
population of lens systems as they are unbiased by extinction in the
lens.

MG0414+0534 was discovered in a survey for lenses conducted at the
VLA, and initial spectroscopy of this object marked it as unusual
because of the extremely red optical continuum ($F_\nu \propto
\nu^{-9}$) (Hewitt et al. 1992).  It has a four-image geometry
characteristic of lensing by an elliptical potential.  The lens galaxy
was first detected in $I$-band by Schechter \& Moore (1993). Subsequently, IR
images were obtained, and a redshift of 2.639 for the source was
determined using near-IR spectroscopy (Lawrence et al. 1995). These IR
images were not of sufficient resolution to detect the lens galaxy
or to obtain separate photometry for components A$_1$ and A$_2$.
Much
improved optical photometry of the four images and lens galaxy and a
detection of lensed arcs have been made with Hubble Space Telescope
(Falco, Leh\'ar, \& Shapiro 1997, hereafter FLS97).  To date, no
redshift has been determined for the lens galaxy.  Because MG0414 is
unusual in its redness, we have pursued further observations to
understand the nature of the unusual quasar colors and their relation
to the lens galaxy.

In this paper we present images of the system, which allow near IR
photometry of all four images and the lens galaxy for the first time.
We then present $K$-band spectra of the components of MG0414 which 
allow us to set limits on the amount of microlensing present.  We conclude
with implications that the galaxy colors have for the presence of dust in
this and other galaxies.

\section{Imaging}

Infrared images at $H$ (1.65$\mu$m) and $K$ (2.2$\mu$m) were obtained
at the Multiple Mirror Telescope (MMT) using a 128$\times$128
NICMOS2 array.  $K$ images were taken on UT 1992 December 7. $H$ and
$K$ images were taken on 1993 March 10.  One of the complications of
observing at the MMT is the need to keep the images from the six
telescopes coaligned.  This drives the observing strategy, which is to
stack the images on a bright star using the IR array, then to take 
three 60s exposures on source, and then to restack.  Before and
after stacking, a 1s exposure of the bright star is recorded and is
used as a photometric and PSF reference.  The seeing was typically
0\farcs7 with 0\farcs2/pixel.

\subsection {PSF Fitting}
Because the resolution of the images is not much greater than the
separations of the components, the only way to determine the
brightnesses of the four quasar components and the lens galaxy is
though PSF fitting.  We developed our own code that fits a model
convolved with a PSF image.  In this case the model consists of four
delta-functions and a deVaucouleurs profile ( $I(r) \propto
exp(-(r/r_e)^{1/4})$).  The general case has 18 free parameters (x, y,
and intensity for all 5 sources, plus the scale length ($r_e$), the
ellipticity, and the position angle for the galaxy).  We are able to
reduce the number of free parameters to 9 by adopting HST values
(FLS97) for the quasar locations and the galaxy location, ellipticity
and position angle. We fit all 5 intensities, but reduce the positional
data to solving for the x and y offsets, scale, and rotation of our
coordinate system relative to the HST data.

Ideally, for a PSF, we would like to use a stellar image from the same
frames that the lens data came from.  Unfortunately this is not possible as
there are no suitably bright sources in the 26\arcsec\ field.  Instead we use
the bright star that was observed before and after each realignment of the
six telescopes.  Because telescope tracking errors are underestimated
in the short exposures, the resulting PSF images are slightly too narrow
compared with the true PSF.  However, the amount of light in the wings
of the PSF should be correct.  To correct for this error, we additionally
convolve the PSF image with a small gaussian profile.  The sigma of this
gaussian is typically 0\farcs1, and is a 10th free parameter of the fit.

\subsection{Results}
Figure~\ref{fig-kmarch} shows the $K$ image from 1993 March.  Panels
A-C show the observed image, model image, and the residual image,
respectively.  The final panel shows the lens galaxy, resulting from
subtracting the four model QSO images only.  In
Table~\ref{tab-photometry} we tabulate the derived magnitudes of the
components of the system.  The galaxy magnitude represents the total
magnitude of the deVaucouleurs profile obtained from the fitting
process.  In Table~\ref{tab-colors} we list the colors of the
components.  The $R$ and $I$ data come from FLS97.  For the galaxy,
FLS97 measure the magnitude in a 1\farcs82 diameter aperture.  For a
deVaucouleurs profile with a 1\farcs5 scale length, 35\% of the total
flux is in this size aperture.  We use this correction factor of
1.14 mag to determine the optical-IR colors of the galaxy.  We use the
mean of the 1992 December and 1993 March $K$ data sets in computing the
$K$ colors.  A direct aperture measurement from
Figure~\ref{fig-kmarch}d yields the same galaxy magnitude to within
10\%.  The Galactic extinction tabulated in NED
\footnote{The NASA/IPAC Extragalactic Database (NED) is operated by the Jet 
Propulsion Laboratory, California Institute of Technology, under
contract with the National Aeronautics and Space Administration.} for
this line of sight is $A_B$=0.48.  This corresponds to $A_K = 0.04$ and
$E(\hbox{$R\!-\!K$})$ = 0.23 (Savage \& Mathis 1979).  We ignore these comparatively
small effects in the following discussions.

Uncertainties for the photometry have been estimated by making a set
of ten simulated images.  Each of these images uses the derived model
image obtained from fitting the original data. Gaussian random
noise is then added to match the noise level in the original data.  We
determine the random uncertainty in the fitted magnitudes by running
the fit on each of the simulated images and computing the variance of
the derived magnitudes.  To allow for errors introduced by PSF
variations, in the $\sigma$ column in Table~\ref{tab-photometry} we
quote the larger of the uncertainties from the simulation and from
taking half the difference between the two $K$ data sets. The
uncertainties for components A$_1$ and A$_2$ are larger than for B
because there is a substantial cross-correlation between the fluxes of
the two close components.  The noise properties of the $H$ and $K$
images are similar so we adopt the same uncertainties for the $H$
magnitudes.

We have performed a number of checks to verify that our fitting
procedure and PSFs are robust.  To verify that the A$_1$/A$_2$ ratio
is not affected by PSF errors we extract image B from the observed
image by subtracting out the other model components.  This image is
used then as the PSF and the data is refit.  The resulting A$_1$/A$_2$
ratio changes by less than 10\% in all three data sets.  Because the
noise in the ``B'' PSF image is quite large, it is not useful for
estimating the flux in image C or the galaxy.  We also have
investigated the effect of changing the scale length of the galaxy,
since FLS97 quote an error bar of roughly 50\% in this parameter.  If
we use $r_e = 0\farcs5$, instead of 1\farcs5, ( a more drastic change
than is required by the errors), we find that the galaxy flux measured
in the 1\farcs8 diameter aperture decreases by only 0.08 mag. The total
flux integrated over the deVaucouleurs profile does drop by a factor of
two.  The fluxes in the point sources change by roughly $2\%$ each.
Thus the derived point source photometry and the galaxy colors are
independent of the assumed galaxy scale length.

Figure~\ref{fig-qsocol} shows a color-color plot for the four images
in MG0414.  Also drawn is a reddening vector, showing that the
differing colors can be naturally explained by variable amounts of
extinction along the line of sight to the quasar.   The length of the
line corresponds to $\Delta E(\bv) = 1$.  The dusty galaxy
explanation was previously argued by Lawrence et al. (1995) and is
strengthened here with the separate A$_1$ and A$_2$ measurements at $K$.

\section{Spectroscopy and limits on microlensing}
If microlensing were taking place, we would expect
that the continuum emission, emanating from the blackhole accretion disk,
would be affected, but that the H$\alpha$ line, coming from the larger, 
parsec-sized, broad-line region would  be unaffected by microlensing.
In other words, microlensing will change the H$\alpha$ equivalent width.
Because microlensing affects each of the images independently, we would
expect the equivalent widths of the four components to have different
values if microlensing were significant.  Thus a single comparison of 
of the equivalent widths of the individual component is sufficient to
set a limit on the amount of microlensing; no time series is required.

A $K$-band spectrum was obtained at the MMT on UT 1993 November 25
with FSPEC (Williams et al. 1993) at a resolution of 800.  The slit
location is shown in Figure~\ref{fig-slit}.
The spatial FWHM was 1\farcs1 with 0\farcs4 pixels.  The spatial
resolution was good enough that we could extract two separate spectra
for components A$_1$/A$_2$ and B.  Figure~\ref{fig-spectrum}(a) shows
the spectrum of the A components, and (b) shows the ratio of the
A and B spectra.  The spectrum is dominated by a single emission line,
identified as H$\alpha$ at a redshift of 2.639 (Lawrence et al. 1995).
We compute that the equivalent width ratio between the two components is
$W_B / W_A = 0.99 \pm 0.03$.

Though we cannot directly measure the line/continuum ratio between
A$_1$ and A$_2$, we can obtain a limit for the close pair by measuring
the change in spatial position of the spectrum as a function of
wavelength.  If the continuum is enhanced in image A$_1$, the centroid
of the line will be biased towards A$_1$ in the continuum region
relative to the location in the line.  Figure~\ref{fig-spectrum}(c)
shows a plot of the spatial position of the A$_1$/A$_2$ spectrum as
a function of wavelength.  The centroid in the line (between the 
dashed lines) compared with the surrounding continuum differs
by $0\farcs003\pm0\farcs017$.  Using 0\farcs017 as an upper limit, 
we derive an upper limit to the flux change between A$_1$ and A$_2$
of 0.22.  This is not as good a limit as for  A vs. B, but could
be improved with a higher S/N ratio spectrum.  The technique could
be useful for monitoring for microlensing in lenses with very close
separations.

Witt, Mao, \& Schechter (1995) (hereafter WMS95), calculated
histograms of the expected flux changes in the images of MG0414
due to microlensing for
various source sizes sampled over time.  They assumed the color
differences among the components of MG0414 were due to microlensing
and then derived an upper limit for the size of the continuum emission
region.  We can use the same distributions, combined with our upper
limit on the microlensing effect on the line/continuum ratio, to
derive a lower limit to the source size.  Note that it is is not
appropriate to consider the WMS95 lower limit and our upper
limit as bracketing values for the correct source size as they
originate from conflicting assumptions as to the amount of
microlensing detected.

For a source size of $10^{16} {\rm cm} \times (\langle M \rangle / 0.1
M_{\sun})^{1/2}$, where $\langle M \rangle$ is the average stellar
mass of the lens galaxy, WMS95 predict the the A$_1$/A$_2$ continuum
ratio will have an RMS scatter of 0.47 mag.  The probability of
observing $\Delta m_{A_1,A_2} < 0.22$ is 0.33.  If we take their
A$_1$/A$_2$ distribution as typical for A/B, then the probability of
observing $\Delta m_{A,B} < 0.03 $ is 0.05.  This is an upper limit to
the probability because the fluctuations in B are larger than those in
$\rm A_1$ and $\rm A_2$. The joint probability of observing both lower
limits is 0.015.  Thus, it is likely that either the source size is
larger than stated, or the assumptions about the stellar population in
the lens galaxy is incorrect.  The microlensing models used in WMS95
assume that all the galaxy mass is in stars, and for images A$_1$ and
A$_2$, the assumed mass density corresponds to $4.5\times 10^{10}
M_{\sun} \rm arcsec^{-2}$, assuming $z_g = 0.5$, $H_0 = 50$, and
$q_0=0.5$.  Combining the galaxy profile in FLS97 with the observed
$I-K$ colors and the predictions of the stellar mass-light ratio from
the Bruzual \& Charlot (1993) models for an elliptical galaxy, we
estimate the stellar density to be $1.4\times10^{10} M_{\sun} \rm
arcsec^{-2}$, at the location of A$_1$/A$_2$.  This model assumes a
Salpeter IMF with stars ranging in mass from 0.1 to 125 $M_{\sun}$.
A resolution of the discrepancy between the predicted microlensing
probability and the apparent lack of microlensing will require more
detailed modeling, using the observed galaxy profile and knowledge 
of low-mass stellar populations from Galactic microlensing experiments.

\section{Discussion}

\subsection{Different colors for the components} Based on
Figure~\ref{fig-qsocol} we have argued that the different colors for the
different components are due to differential extinction along the
lines of sight and not likely due to microlensing.
We can also rule out two other mechanisms which might
cause apparent differing colors.  If the source were varying
intrinsically, we could infer different colors for the components
since the observations were made at different times.  However, the
good agreement in the $I$-band photometry between Schechter \& Moore
(1993) and FLS97 make fluctuations at the required 0.5 magnitude level
extremely unlikely.  The second possible mechanism for producing
differential colors might be if the source size or geometry were a
function of wavelength.  This too can be excluded, via a simple model.
If we consider the A$_1$/A$_2$ ratio as a function of source position, we
find that for those source positions which produce images within 0\farcs2
of the observed image locations, the flux ratio changes by less than 5\%.
The $K$ and $R$ positions agree to much better than this; the flux
ratios differ by much more. Therefore source geometry can be
excluded as a mechanism for the differential image colors.

\subsection{Estimating the lens redshift}
We can set a lower limit for the lens galaxy redshift using the fact
that the (unmeasured) velocity dispersion (or equivalently the mass)
of the lens galaxy is related both to the image separation and the
galaxy luminosity as a function of redshift.  Redshifts where the
velocity dispersion derived from the the image separation matches that
derived from the galaxy photometry are viable lens redshifts. The
Faber-Jackson relation (Faber \& Jackson 1976) relates a galaxy's
luminosity to its velocity dispersion.  The  relation we
use, $\log \sigma = -0.114 B + 3.96$ at $z$=0.023, is derived from
$B$-band photometry for the Coma cluster (Dressler et al. 1987). We
transform the relation from rest-wavelength $B$ to the observed
$K$-band using the passively evolving spectral energy distribution for
an elliptical galaxy described by McLeod \& Rieke (1995).  This model
was generated from the GISSEL galaxy evolution code (Bruzual \&
Charlot 1993).  We prefer to use the observed $K$ magnitude rather
than $R$ or $I$ for two reasons. First, at high redshift, the K
corrections are better known than at shorter wavelengths, and second,
if dust is responsible for the redness of the galaxy, the $K$
magnitude will better represent the true stellar luminosity of the
galaxy. The velocity dispersion of the galaxy, $\sigma$, is related to
the lens image separation, $\beta$, via $\beta = 4\pi(\sigma/c)^2
D_{ls}/ D_s$, where $D_{ls}$ and $D_s$ are the distances from the lens
and the observer, respectively, to the source (e.g.  Blandford \&
Narayan 1992).  Figure~\ref{fig-fj} shows a plot of the expected
velocity dispersion as a function of the lens redshift derived from
the lens model (dashed) and the Faber-Jackson relation (solid).  The
light lines represent 1$\sigma$ errors due to the observed scatter in
the Faber-Jackson relation.  The point where the two curves cross is
the best solution for the lens galaxy redshift.  We can see that the
relation is degenerate at high redshift so we can set only a lower
$1\sigma$ limit of z=0.45.

We can set an upper limit for the lens redshift using the observed $K$
magnitude of the lens galaxy and assuming a bright limit for the
absolute magnitude.  There are a few known radio galaxies with
$K\sim16$ with redshifts between 0.75 and 1.0 (McCarthy 1995).
However, these are very rare objects and the lens galaxy is not a
radio galaxy.  A typical brightest cluster galaxy has $L=4L_*$
(Sandage 1972), i.e., $M_K = -24.6 + 5 \log h$ (Gardner et al. 1997).
Such a galaxy would have the observed brightness of $K$=16 
if it were at z=0.6.  Because there is no apparent cluster
surrounding MG0414 (FLS97), we view z=0.6 as a reasonable upper limit
on the lens galaxy redshift.

\subsection{Redness of the Galaxy}
The lens galaxy is as red as any known galaxy of comparable magnitude.
Figure~\ref{fig-ik} shows a histogram of $I$-$K$ colors of field galaxies
with $15.65<K<16.65$ from the Hawaii field galaxy surveys (Gardner 1995).
The arrow marks the location of the MG0414 lens galaxy.  A similar result
can be seen for $R-I$ (Tyson 1988).   

What could lead to such red colors?  There could be at least two
potential explanations: 1) The lens galaxy is at redshift z$\sim$1.
In this case the K-corrections for a normal galaxy would give it
approximately the correct colors.  However, this would require the
galaxy to be extremely luminous and we have argued above that the
redshift is likely less than 0.6.  2) The galaxy is dusty (Lawrence et
al. 1995).  We prefer this explanation as it simply explains the
unusual colors of both the galaxy and the quasar images
simultaneously.

Figure~\ref{fig-galcol} shows the observed $R-I$ and $I-K$ colors of
the lens galaxy.  The solid line is the expected track for the model
elliptical galaxy as a function of redshift, showing that the observed
colors are too red in both axes.  The dashed line shows the effect of
dereddening the galaxy assuming a uniform mixture of stars and dust.
Though the assumption of a uniform mixture is probably not
quantitatively correct, it does qualitatively produce reasonable
unreddened colors.  This figure shows clearly the difference between
redshift and dust reddening.  We can see that the galaxy colors give
z=0.6 and $A_V=7$.  The redshift is consistent with the upper limit
derived above.  The reddening matches reasonably well with Lawrence et
al.(1995) who argued that A$_V\sim5.5$ in the galaxy is required to
produce the observed QSO colors, assuming a normal unreddened QSO
spectrum.

Another extremely red lens is MG1131+0456, which has been studied in
the IR by Larkin et al. (1994).  They derive $R-K$=5.3 for the lens
galaxy, which is 0.5 mag bluer than MG0414 even after accounting for
Galactic extinction. They conclude that the lens galaxy of MG1131 is
consistent with the spectrum of NGC4889 redshifted to 0.75, but that
the quasar colors require several magnitudes of $A_V$.  Our elliptical
evolution model predicts $R-K$=4.5 at z=0.75, leading us to conclude
that reddening in the MG1131 lens is also consistent with the galaxy
colors.  We note that although the NGC4889 colors are 0.4 mag redder
in $R-K$ than our model, we would still argue that some reddening is
required in MG1131 to match the NGC4889 colors.

The colors of the MG0414 galaxy place it at the end of a red tail of
the distribution of the colors of galaxies with similar brightness.
In addition to the observed $I-K$ colors of field galaxies,
Figure~\ref{fig-ik} shows the expected distribution of colors for
$15.65<K<16.65$ galaxies using the evolving elliptical SED.  A
significant number of galaxies lie redward of the model, including the
MG0414 lens galaxy.  Figure~\ref{fig-ik} is qualitatively consistent
with the analysis of Lawrence et al (1995) which showed that two out
of six radio selected lens systems showed large amounts of extinction.
The large amounts of extinction in MG0414 suggest that the presence of
dust in galaxies at z$\sim$0.5 may be a dominant cause of the colors
being redder than predicted by the models.

\section{Conclusions}
Near IR imaging and spectroscopy of MG0414 support the hypothesis that
dust in the lens galaxy is responsible for the red colors of the QSO
images.  The colors of the lens galaxy itself are consistent with
large amounts of internal extinction.  $K$-selected field galaxy
surveys contain a significant number of galaxies that are redder than
are predicted by stellar evolution models.  The extinction seen
through the red MG0414 lens galaxy suggests that dust may be the cause
of red field galaxies.  The lack of variation in the line/continuum
ratios among the components argue against significant amounts of
microlensing.  This in turn argues against a small size for the QSO
accretion disk, or for problems in the stellar population assumptions
used in the microlensing models.

\acknowledgments
We thank E. Falco for providing data in advance of publication.

\newpage

\begin{deluxetable}{ccccc}
\tablenum{1}
\tablewidth{0pt}
\tablecaption{Photometry of components\label{tab-photometry}}

\tablehead{
\colhead{} &\colhead{ 1992 Dec } & \multicolumn{2}{c}{ 1993 March} & \\
 \cline{3-4} \\
\colhead{Component}&\colhead{$K$}& \colhead{$K$}& \colhead{$H$} &\colhead{$\sigma$}
}
\startdata
A$_1$ & 14.33 & 14.34 & 15.31 &0.05\\
A$_2$ & 14.53 & 14.56 & 15.51 &0.05\\
B     & 15.36 & 15.36 & 16.28 &0.01\\
C     & 16.19 & 16.24 & 17.16 &0.02\\
G     & 16.20 & 15.89 & 16.79 &0.15\\
\enddata
\end{deluxetable}

\begin{deluxetable}{cccc}
\tablenum{2}
\tablewidth{0pt}
\tablecaption{Colors of components\label{tab-colors}}

\tablehead{
\colhead{Component} & \colhead{$H-K$} & \colhead{$R-K$} &\colhead{$R-I$} 
}
\startdata
A$_1$ & 0.98 & 8.42 & 2.17 \\
A$_2$ & 0.96 & 9.21 & 2.35 \\
B     & 0.92 & 8.13 & 2.13 \\
C     & 0.94 & 8.04 & 2.05 \\
G     & 0.74 & 6.04 & 1.81 \\
\enddata
\end{deluxetable}

\newpage

\plotone{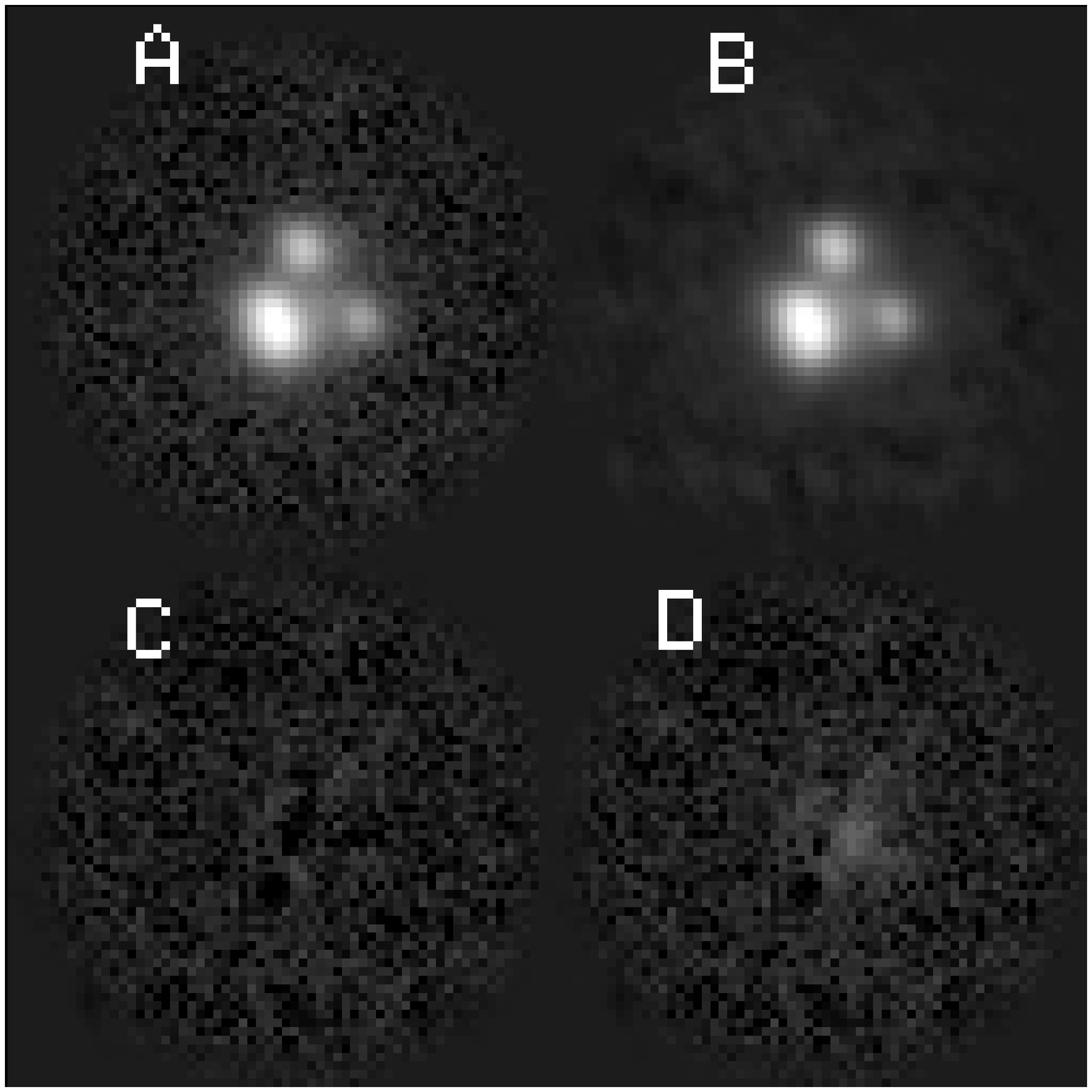}

\figcaption[McLeod.fig1.ps]{\label{fig-kmarch} $K$-band image of MG0414 obtained
1993 March. The four panels show a) the observed image, b) the model image
c) the residual image, and d) the lens galaxy, obtained by subtracting
the four model QSO images.}

\plotone{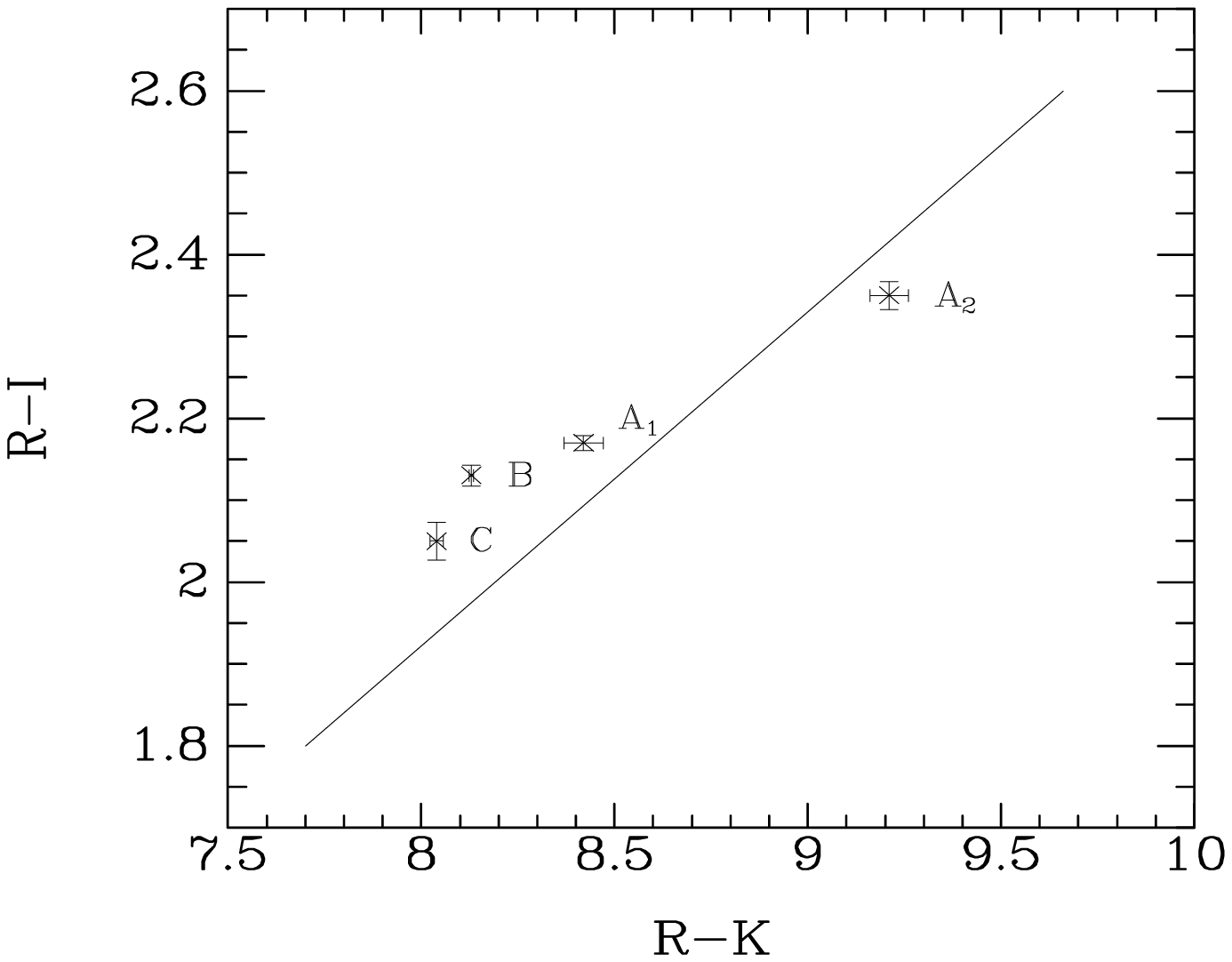}

\figcaption[McLeod.fig2.ps]{\label{fig-qsocol}
The colors of the individual quasar images in MG0414.  The line
represents a reddening law with a length corresponding to a change in
E(B-V) of 1.0.  
}

\plotone{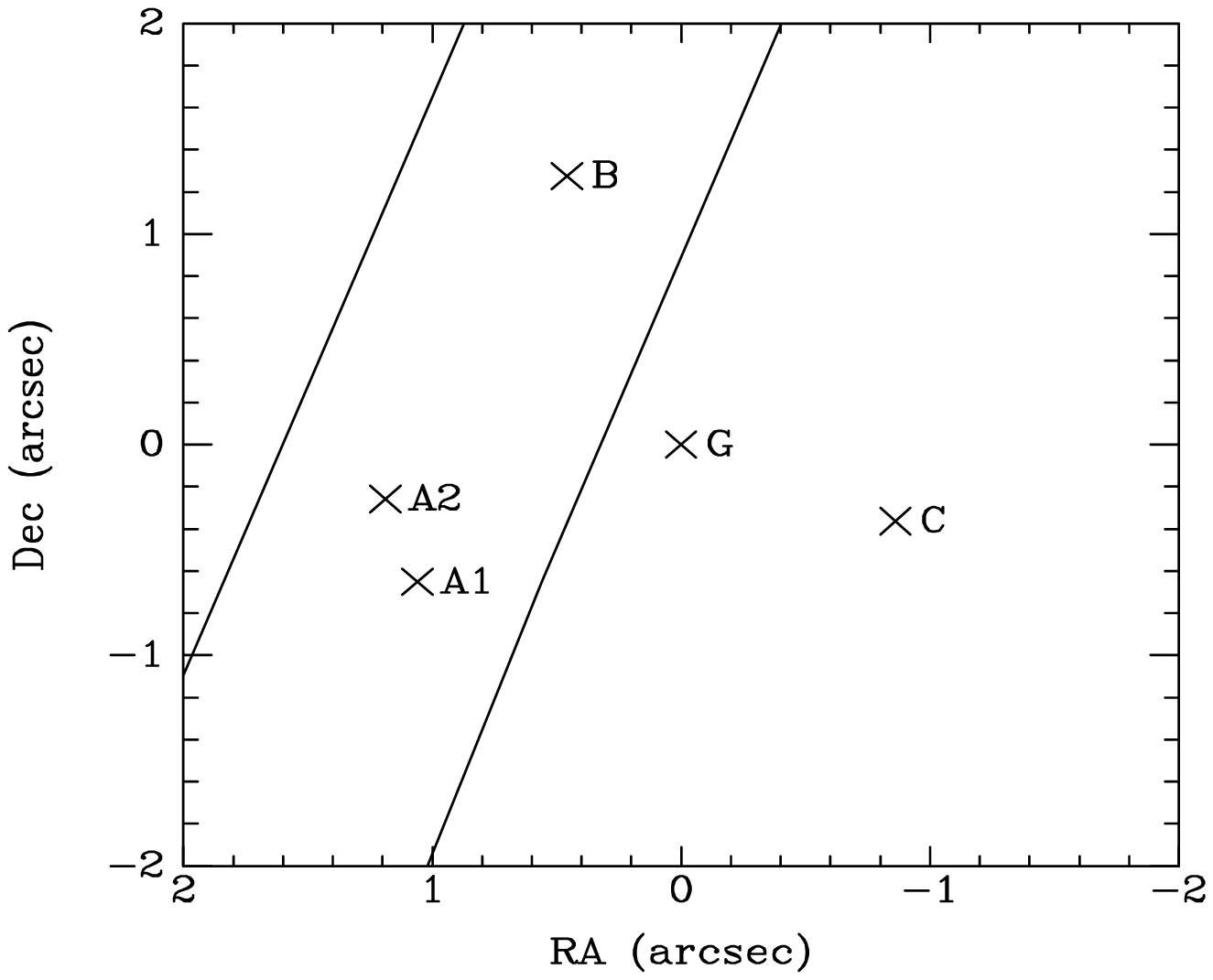}

\figcaption[McLeod.fig3.ps]{\label{fig-slit}
The slit orientation for the $K$-band spectrum.
The 1\farcs2 wide slit covered components A$_1$, A$_2$, and B.
}

\plotone{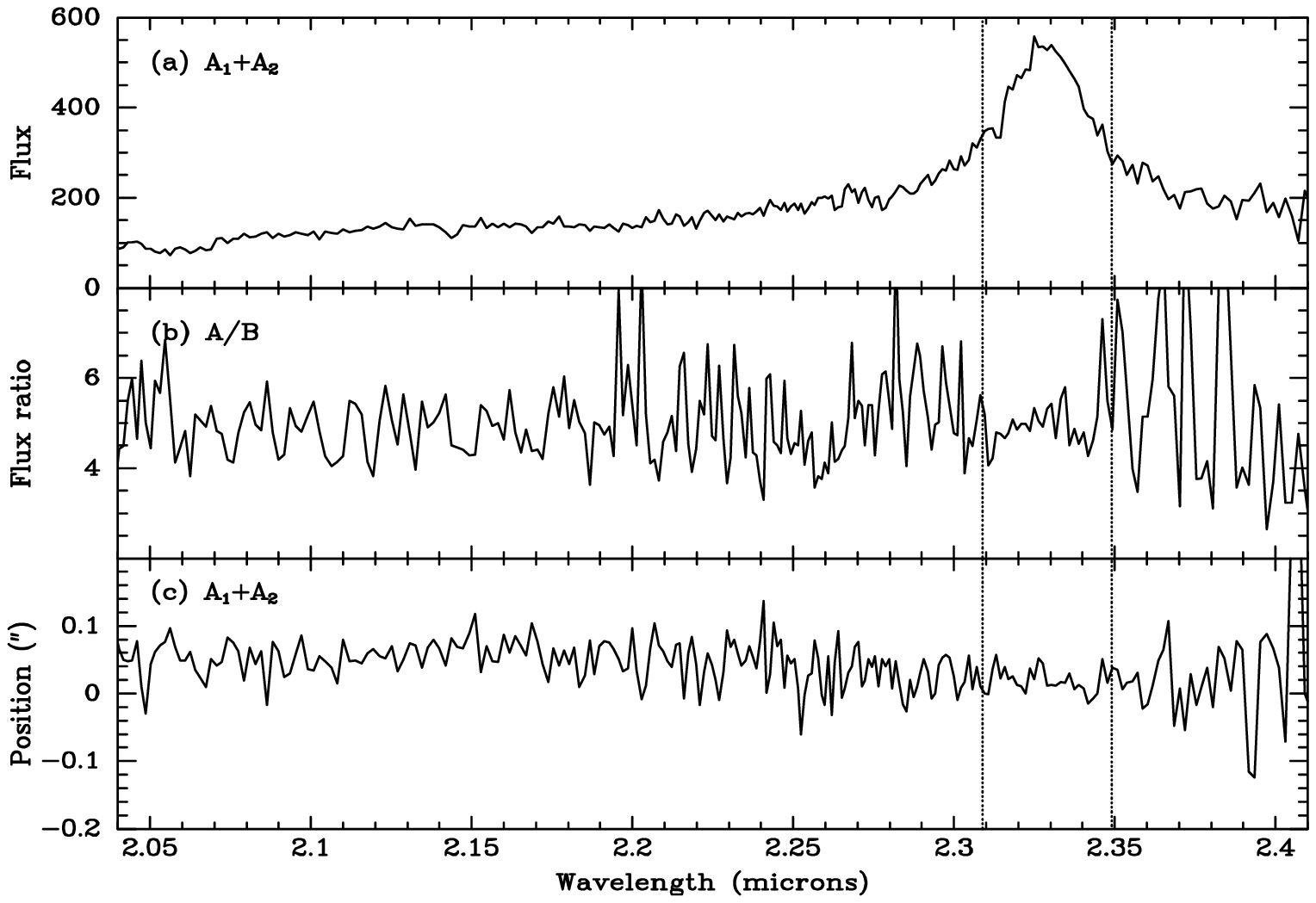}

\figcaption[McLeod.fig4.ps]{\label{fig-spectrum}
a) K-band spectrum of images A$_1$/A$_2$.  The broad emission line
is H$\alpha$ at z=2.36.  b) Ratio of the spectra for images A$_1$/A$_2$ and
B.  The line and continuum ratios are the same, setting an upper limit
on the amount of microlensing.
c) The position of the  A$_1$/A$_2$ spectrum.  The position is the
same in the line and surrounding continuum, setting an upper limit on the
amount of microlensing.
}

\plotone{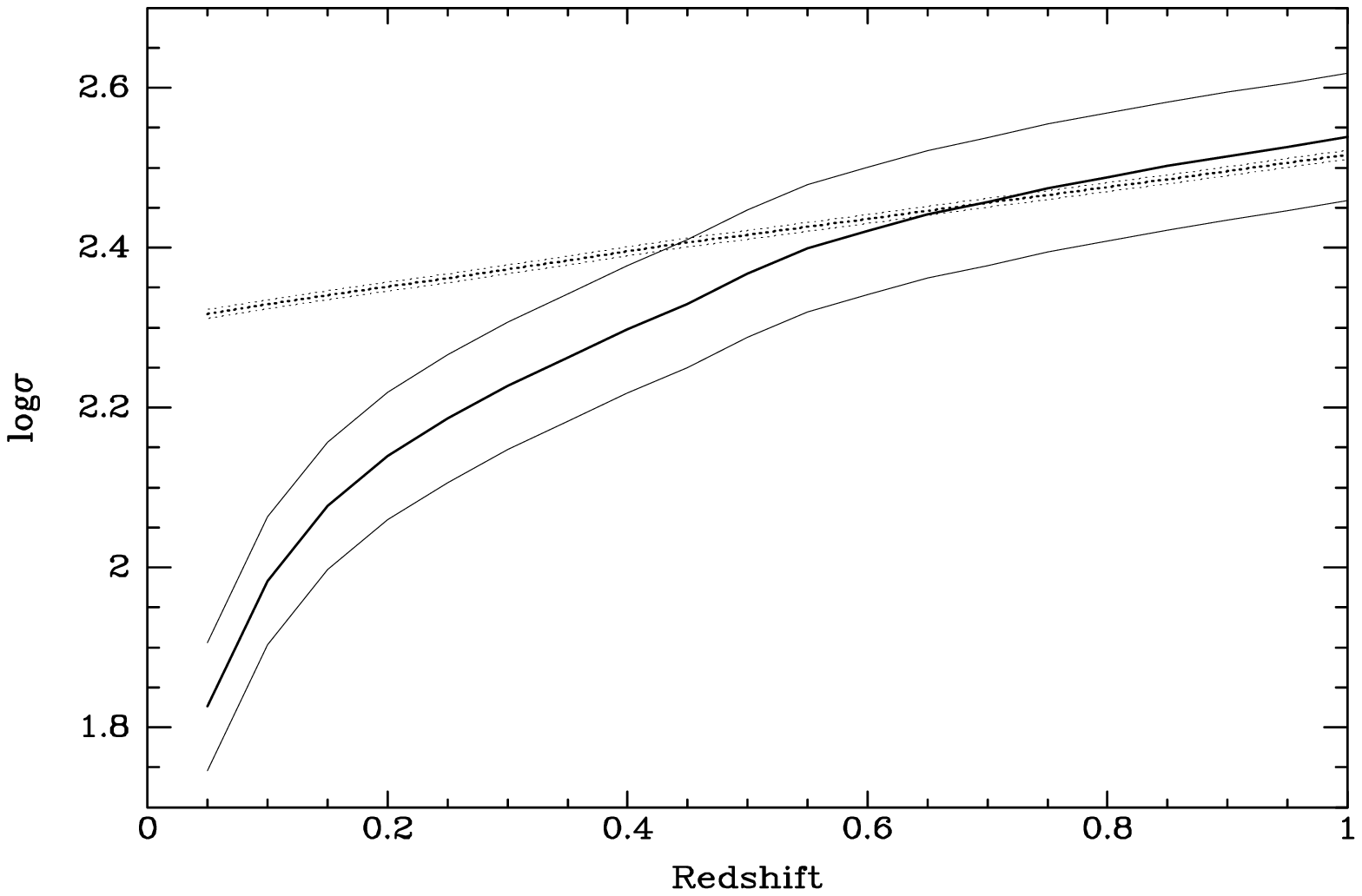}

\figcaption[McLeod.fig5.ps]{\label{fig-fj}
Estimate of the lens galaxy redshift from the Faber-Jackson relation.
The dark solid line shows the derived velocity dispersion based on the
apparent galaxy magnitude as a function of redshift. The thin solid
lines show the 1$\sigma$ errors.  The dashed line represent the
velocity dispersion based on the lens image separation.  We set a
1$\sigma$ lower limit on the lens redshift of 0.45.
}

\plotone{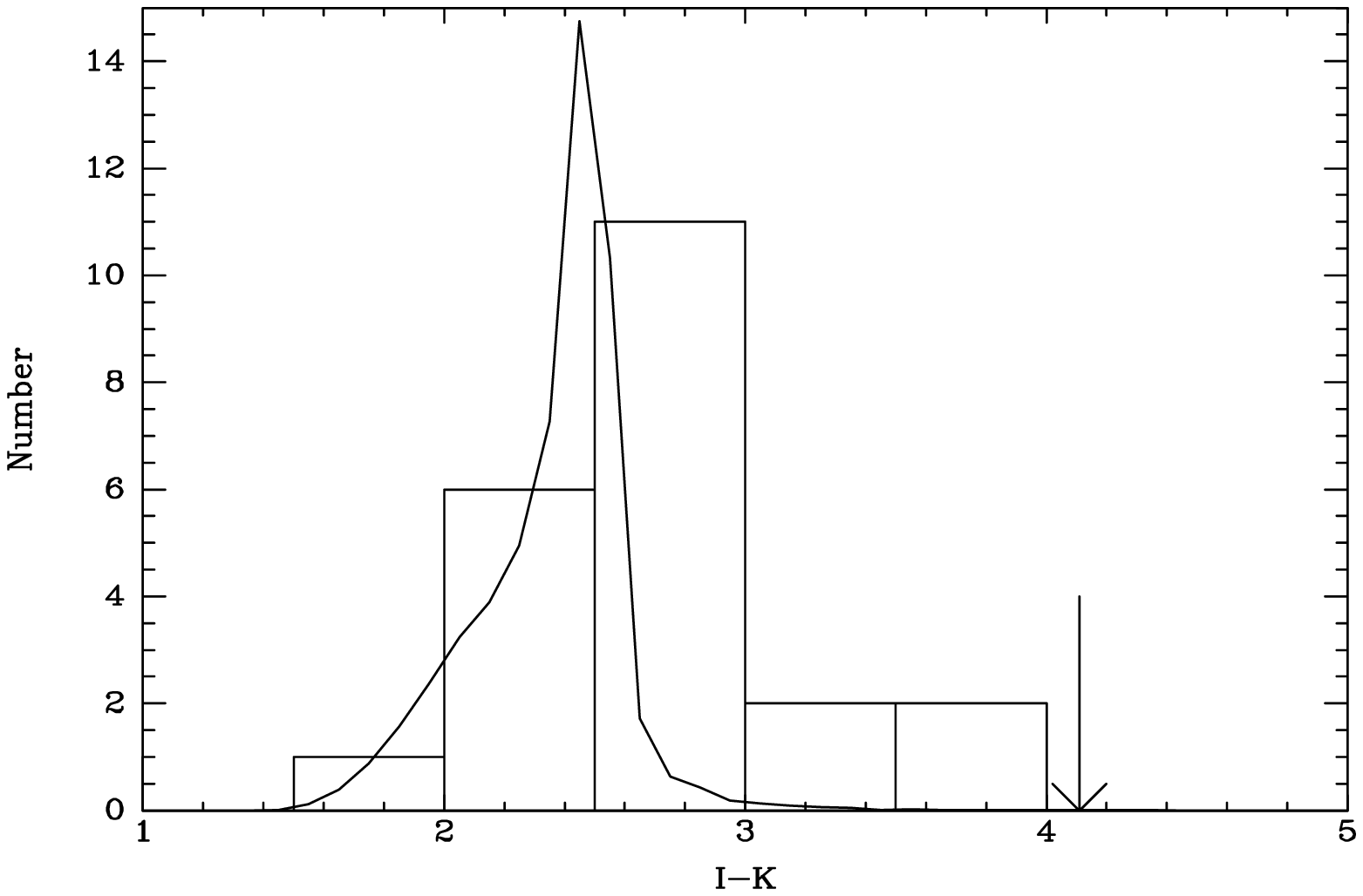}

\figcaption[McLeod.fig6.ps]{\label{fig-ik} The box histogram shows the $I-K$ color
distribution for galaxies $15.65<K<16.65$ taken from Gardner (1995).
The arrow marks the location of the MG0414 lens galaxy.  The curve
shows the expected distribution based on a passively evolving model
(McLeod et al. 1995).}

\plotone{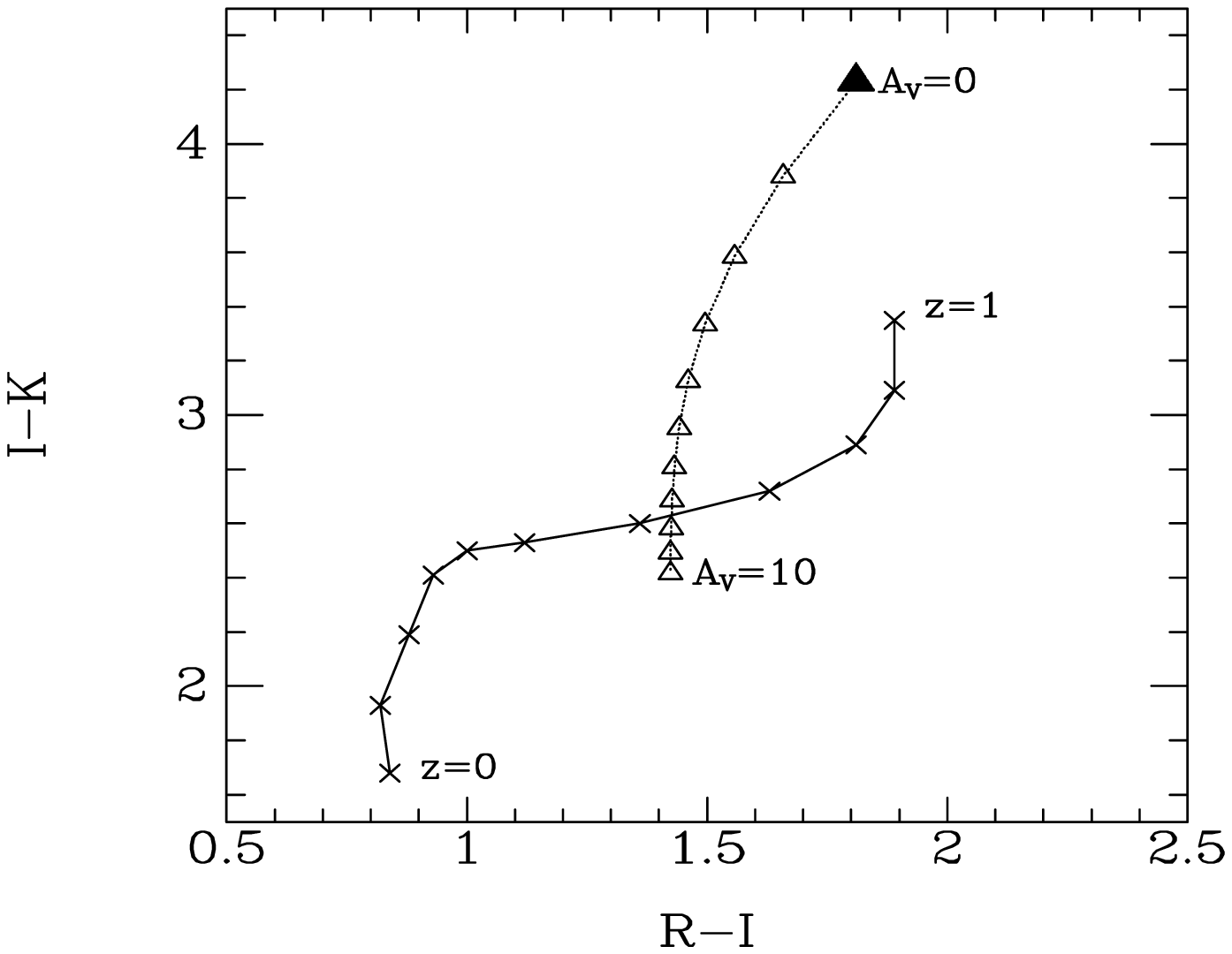}

\figcaption[McLeod.fig7.ps]{\label{fig-galcol}
The solid triangle marks the observed color of the lens galaxy.
The solid curve shows the expected colors for a passively evolving
elliptical galaxy with no dust.  The dotted line shows the effects of
dereddening the lens galaxy assuming a uniform mixture of stars and
dust.  The data are consistent with a heavily reddened galaxy
at z=0.6.
}

\end{document}